\documentclass[aps,a4paper,10pt,twocolumn,nofootinbib]{revtex4} 
\usepackage{amsfonts}
\usepackage{amssymb}
\usepackage{mathrsfs}
\usepackage[mathscr]{euscript}
\usepackage[dvips]{graphicx}
\usepackage[hypertex=true]{hyperref}
\usepackage{fancyhdr}
\usepackage{colordvi}
\usepackage{subfigure}

\newcommand{\XDOI}[1]{\href{http://dx.doi.org/#1}{doi:#1}}

\begin{document}
\title{Carrier-wave steepened pulses and gradient-gated
 high-order harmonic generation using linear ramp waveforms}

\author{S. B. P. Radnor}
\affiliation{
  Blackett Laboratory, Imperial College London,
  Prince Consort Road,
  London SW7 2AZ,
  United Kingdom.}
\author{L. E. Chipperfield}
\email{luke.chipperfield@imperial.ac.uk}
\affiliation{
  Blackett Laboratory, Imperial College London,
  Prince Consort Road,
  London SW7 2AZ,
  United Kingdom.}
\author{P. Kinsler}
\email{dr.paul.kinsler@physics.org}
\affiliation{
  Blackett Laboratory, Imperial College London,
  Prince Consort Road,
  London SW7 2AZ,
  United Kingdom.}
\author{G. H. C. New}
\affiliation{
  Blackett Laboratory, Imperial College London,
  Prince Consort Road,
  London SW7 2AZ,
  United Kingdom.}

\begin{abstract}

We show how to optimize the process of high-harmonic generation (HHG) 
 by gating the interaction using the field gradient of a driving pulse
 with a linear ramp waveform.
Since maximized field gradients are efficiently generated by 
 self-steepening processes, 
 we first present a generalized theory of 
 optical carrier-wave self-steepened (CSS) pulses.
This goes beyond existing treatments,
 which only consider third-order nonlinearity, 
 and has the advantage of describing pulses whose wave forms have 
 a range of symmetry properties.
Although a fertile field for theoretical work, 
 CSS pulses are difficult to realize experimentally
 because of the deleterious effect of dispersion.   
We therefore consider synthesizing CSS-like profiles
 using a suitably phased sub-set of the harmonics present 
 in a true CSS wave form.  
Using standard theoretical models of HHG,
 we show that the presence of gradient-maximized regions on the wave forms
 can raise the spectral cut-off and so yield shorter attosecond pulses.
We study how the quality of the attosecond bursts 
 created by spectral filtering
 depends on the number of harmonics included
 in the driving pulse.

\end{abstract}

\lhead{\includegraphics[height=5mm,angle=0]{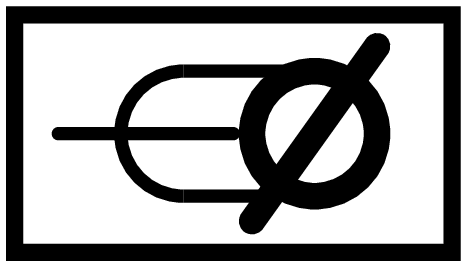}~~CHI2HHG-II}
\chead{~}
\rhead{
\href{mailto:Dr.Paul.Kinsler@physics.org}{Dr.Paul.Kinsler@physics.org}
}
\lfoot{} 
\rfoot{Radnor-CKN-2008} 

\date{\today}
\maketitle
\thispagestyle{fancy}

%
\section{Introduction}

In high-order harmonic generation (HHG), 
 the intensity maxima of a strong driving laser pulse
 tunnel-ionize an atom or molecule.
Once an electron is in the continuum, 
 its acceleration is controlled by the electric field profile $E(t)$.  
The electron is initially driven outwards, 
 but then the field reverses, 
 driving it back for a high-energy recollision with the core,
 approximately three-quarters of a cycle later
 \cite{Schafer-YDK-1993prl, Corkum-1993prl},
 in which an attosecond burst
 of extreme ultraviolet (XUV) radiation is generated.
Indeed,
 each half-cycle of the driving pulse initiates this sequence of events,
 so a train of XUV bursts a few hundred attoseconds long is produced. 
While a train is useful for probing ultrafast dynamics, 
 a primary goal of attosecond science is the generation 
 of single bursts of XUV radiation, 
 and various gating schemes have been employed to achieve this.
One scheme already in use is polarization gating 
 \cite{Budil-SPL-1993pra, Corkum-BI-1994ol, Sola-MECSPVBCSVSN-2006np, 
       Sansone-BCVAFPVAVSDN-2006s}; 
 in this technique,
 the sense of circular polarization
 reverses smoothly through the driving pulse, 
 passing only briefly through linear polarization
 at which point an attosecond burst is generated.

Another technique is to ``gradient-gate'' the HHG interaction, 
 which can be done most simply by an additional
 driving field at double the frequency, 
 suitably phased with respect to
 the fundamental field; 
 this is just the case of two-color pumping
 \cite{Pfeifer-GANL-2006ol, Liu-KSW-2006pra, Oishi-KSKM-2006oe, 
       Merdji-ABCCPJNL-2007ol}.
In the present paper, 
 we push the concept of gradient-gating to its limits by considering 
 how to achieve the steepest gradients possible.
In principle, 
 this can be done with carrier-wave steepening and shocking, 
 a process first identified by Rosen \cite{Rosen-1965pr} in 1965.
He showed that for a instantaneous third-order ($\chi^{(3)}$) nonlinearity, 
 field gradients could become infinite 
 (i.e. shocks could form) after a finite distance of propagation.  
As we demonstrate in Section \ref{S-shocking},
 the theory can be extended
 to include arbitrary combinations of nonlinear terms, 
 which opens up possibilities for controlling the symmetry of the pulse 
 at the same time as generating a gradient-gate.  
We then proceed to discuss the previously unaddressed
 second-order $\chi^{(2)}$ case, 
 which turns out to give wave forms particularly useful for HHG.

Unfortunately 
 it is very difficult to generate carrier-wave self-steepened (CSS) pulses
 because dispersion overwhelms the steepening process
 \cite{Kinsler-RTN-2007pre}.   
In Section \ref{S-practical},
 we therefore discuss the possibility of synthesizing 
 the steepened profiles 
 by combining a suitably phased sub-set of the harmonics
 present in a true CSS profile; 
 optical wave form synthesis of this kind is already 
 exploited in other contexts (see, e.g., \cite{Jiang-HLW-2007np}).
Such a synthesis is in any case  
 the logical extension of both the two-color pumping scheme
 and the gradient-gate concept.

In Section \ref{S-HHG}, 
 we compare simulations of HHG driven by standard pulses
 and by synthesized CSS-like wave forms. 
In the latter case, 
 we demonstrate the importance of the relative phasing
 of the spectral components by comparing the results
 to those based on phase-flattened (i.e. field-maximized) 
 pulses with equivalent spectral content. 
We also show that the efficacy of the synthesized profiles
 for HHG is progressively enhanced
 as the number of harmonics is increased.
High-pass filtering then leads to the production of attosecond XUV bursts, 
 and the presence of even harmonics in $\chi^{(2)}$ CSS-like pulses
 is shown to increase the degree of isolation of the bursts.  
A genetic algorithm is used to show that sawtooth profiles
 are ideal for driving HHG, 
 which is in line with the general conclusions of this study,
 as summarised in Section \ref{S-conclusion}.

%
\section{Carrier-wave self-steepening}\label{S-shocking}

The self-steepening of an optical pulse envelope
 was first studied by DeMartini \emph{et al.} 
 in 1967 \cite{Demartini-TGK-1967pr}, 
 and is a well-known phenomenon associated with self-phase modulation.  
Surprisingly, however, 
 the possibility of shock formation on the optical carrier
 was considered even earlier in a 1965 paper by Rosen \cite{Rosen-1965pr}, 
 who showed that, for a third order ($\chi^{(3)}$) nonlinearity, 
 a field discontinuity (or shock) can develop under certain circumstances
 after a finite distance of propagation.  
Before a carrier shock forms, 
 or before its onset is halted by (e.g.) dispersion
 or the nonlinear response time, 
 the carrier-wave undergoes self-steepening. 
This phenomenon received little attention
 from the optics community for more than 30 years, 
 until it was revisited in the 1990s
 by Flesch \emph{et al.} \cite{Flesch-PM-1996prl,Gilles-MV-1999pre}, 
 who performed finite-difference time-domain (FDTD) simulations
 of the process; 
 more recently, 
 the role of dispersion
 was examined in more detail \cite{Kinsler-RTN-2007pre}.

The theory of carrier shocking and self-steepening is an important 
 way of determining the profiles of pulses with a maximized field gradient, 
 and so is needed when considering the limits of a gradient-gated HHG scheme.
Although a fertile field for theoretical work, 
 CSS pulses are difficult
 to realize experimentally because of the deleterious effect of dispersion.
Nevertheless, 
 the theory provides important information when 
 maximizing the gradient-gating effect in HHG.

%
\subsection{General case}\label{S-shocking-general}

We use the method of characteristics (MOC) \cite{whitham} to predict the 
 shocking distance for a pulse in a dispersionless medium 
 containing an arbitrary combination of nonlinear terms.
The theory is based on the one-dimensional (1D), 
 sourceless, 
 plane-polarized Maxwell's equations for a field
 propagating in the $z$ direction,
 and a material response characterized by the 
 electrical displacement field
~
\begin{eqnarray}
\label{eq:d}
  D
&=&
  \epsilon_{0}\left(E+\chi^{(1)}E+\sum_{m>1}\chi^{(m)}E^{m}\right),
\end{eqnarray}
where $\chi^{(m)}$ refers
 to the $m$-th order nonlinear susceptibility of the medium, 
 which is assumed to be instantaneous. 
This extends the model used by Kinsler \cite{Kinsler-2007josab}, 
 since it allows for combinations of all orders of nonlinearity.
Propagating pulses with a field profile $E(t)$
 using this model leads to self-steepening
 of the carrier-wave.
Eventually, 
 a localized  {\em infinite} field gradient occurs
 on the wave form at a distance
~
\begin{eqnarray}
  \label{eq:moc_n}
  S
&=&
 - 
   \textrm{Min} 
    \left[
      \frac{2c\sqrt{n_{0}^{2}+\displaystyle\sum_{q>1} q\chi^{(q)}E^{q-1}}}
           { \displaystyle\sum_{m>1} m \chi^{(m)}\frac{dE^{m-1}}{dt}}
    \right]
.
\end{eqnarray}
Some exotic features of this prediction 
 are discussed in \cite{Kinsler-2007josab}; 
 a longer derivation is given in appendix \ref{S-MOC}.

To verify the predictions of eqn.~(\ref{eq:moc_n}), 
 we used the 
 pseudospectral spatial domain (PSSD) method \cite{Tyrrell-KN-2005jmo}
 to propagate pulses for various orders of nonlinearity. 
The numerical shocks were detected using
 the local discontinuity detection (LDD) method \cite{Kinsler-RTN-2007pre}, 
 which is useful for shock detection within a discretized system. 
We also used a hybrid propagation technique, 
 combining directional variables \cite{Kinsler-RN-2005pra}
 and a wideband envelope \cite{Genty-KKD-2007oe}, 
 to check the results of the steepened and shocked wave forms. 
In all cases, the LDD (numerical) shocking distances obtained
 from our simulations were in good agreement with those 
 predicted by eqn.~(\ref{eq:moc_n}).

\begin{figure}[htb]
\centering
\includegraphics[angle=-90,width=0.68\columnwidth]{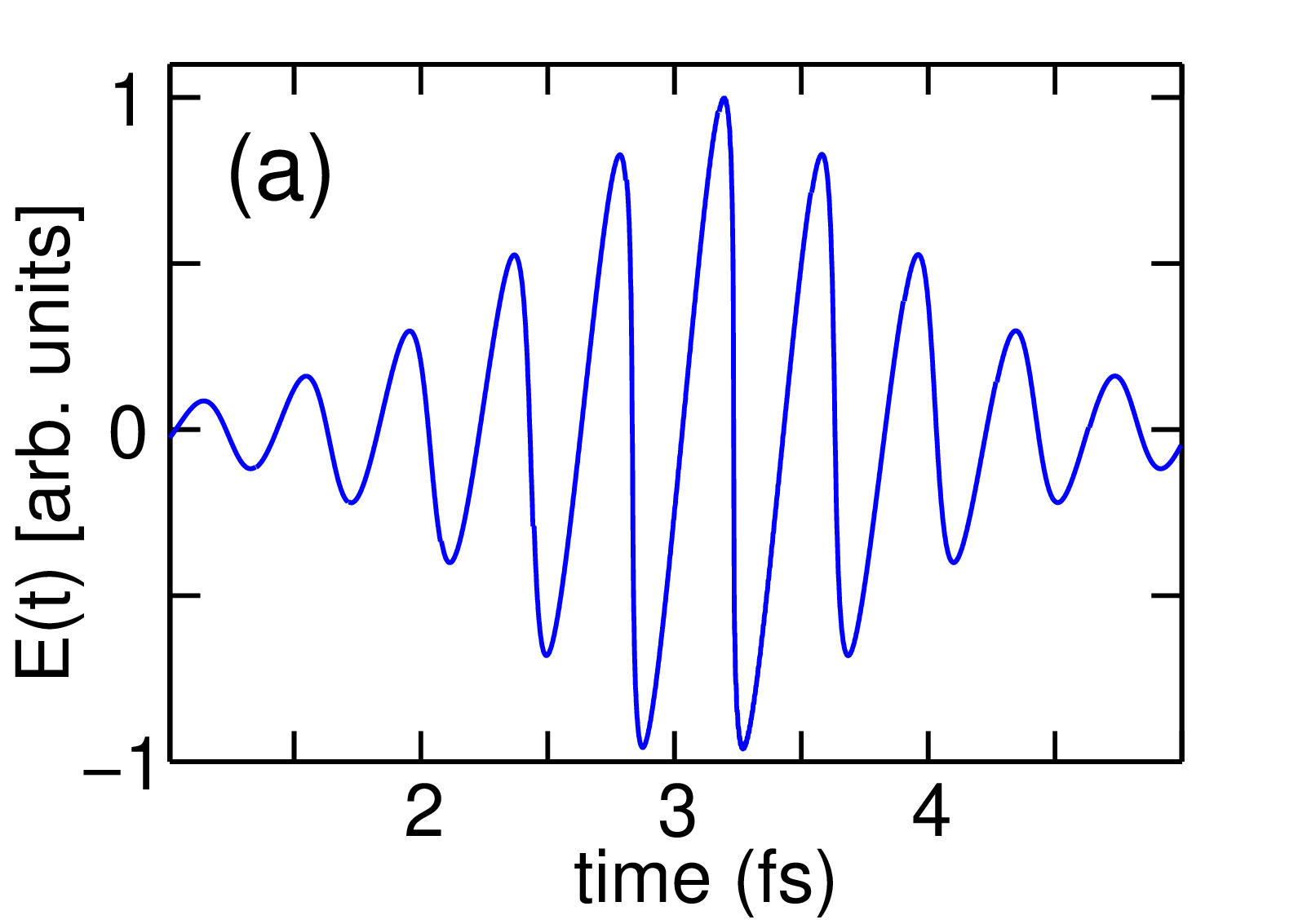}
\includegraphics[angle=-90,width=0.68\columnwidth]{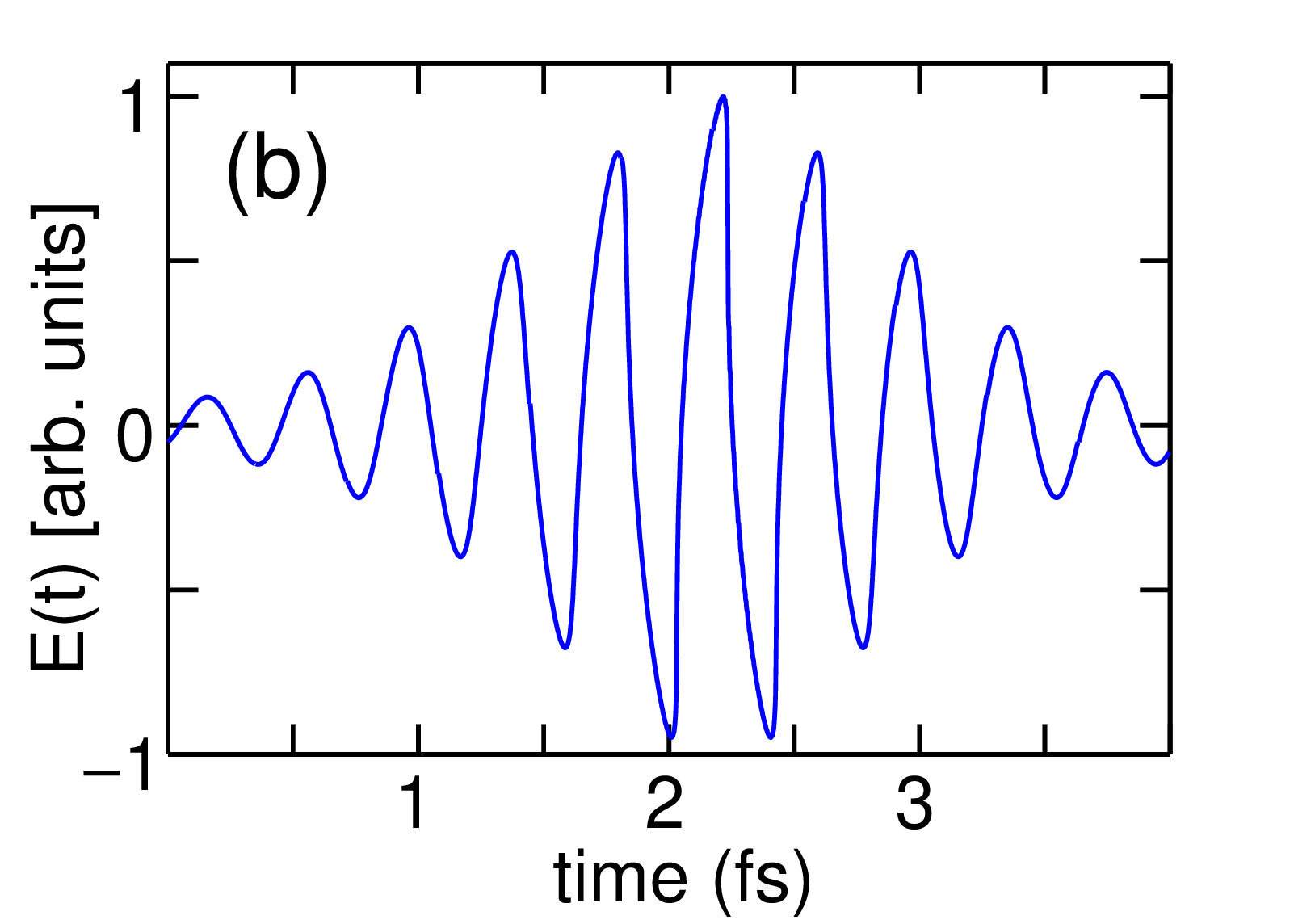}
\includegraphics[angle=-90,width=0.68\columnwidth]{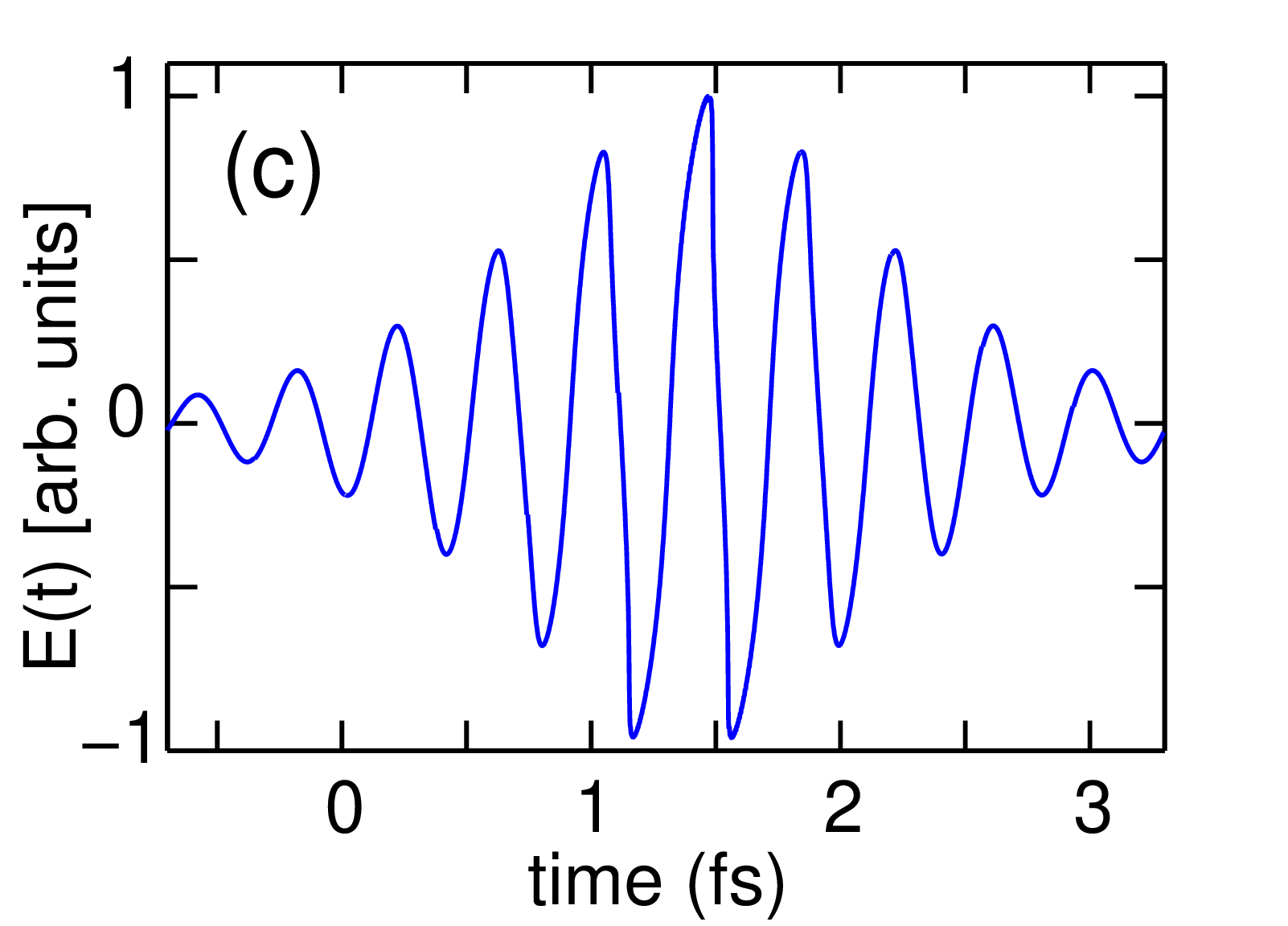}
\caption{
\label{fig:profiles}
Self-steepening at the LDD shocking distance for 
 (a) $\chi^{(2)}$, 
 (b) $\chi^{(3)}$, 
 and 
 (c) $\chi^{(4)}$ nonlinearities.
Although a material with only a $\chi^{(4)}$ nonlinearity is not realistic, 
 its inclusion helps illustrate some important features.  
We can test inversion symmetry by replacing $E$ with $-E$.
The initial pulse has 
 $E = E_0 \sin (\omega t + \phi_1) \mathrm{sech}(0.28\omega t/\tau)$,
 with a wavelength of 800 nm, $\phi_1=0$, and $\tau=1$; 
 $n_0=1$.
The nonlinear strength is $\chi^{(n)} E^{n-1} = 0.02$.
}
\end{figure}

Fig.~\ref{fig:profiles} illustrates typical 
 CSS pulses for low-order nonlinearities, 
 and displays some striking properties.  
In particular, 
 figs.~\ref{fig:profiles} (a) and \ref{fig:profiles} (c)
 (for $\chi^{(2)}$ and $\chi^{(4)}$) 
 both have an underlying sawtooth shape, 
 where the unit of repetition is a whole cycle.  
This is because both even \emph{and} odd harmonics are present
 in the spectrum, 
 and the inversion symmetry of the initial pulse is broken; 
 positive lobes of $E(t)$ lean to later times, 
 and negative lobes to earlier times,
 since (for $\chi^{(2)}$)
 the change in characteristic velocity is proportional to ${dE}/{dt}$, 
 and (for $\chi^{(4)}$) to
 ${dE^3}/{dt}$.
In contrast, 
 fig.~\ref{fig:profiles} (b) (for $\chi^{(3)}$) shows
 both positive and negative lobes 
 leaning to later times, 
 since the change in characteristic velocity 
 is proportional to ${dE^2}/{dt}$.
Because odd nonlinearities produce a cascade
 containing \emph{only} odd harmonics,
 every half cycle is similar
 and the pulses preserve inversion symmetry. 
Another noteworthy feature is that, 
 for higher-order nonlinearities, 
 the steepening becomes increasingly localized to
 within the most intense carrier
 oscillation(s).

%
\subsection{Special case: $\chi^{(2)}$ CSS pulses}\label{S-shocking-chi2}

We now present a brief review of self-steepening and shocking
 for $\chi^{(2)}$ nonlinearities.
We focus on this case 
 because it has not previously been addressed
 and because of the relevance to HHG 
 discussed later in  Section \ref{S-HHG}.
Of course, 
 nonlinearities of higher order than third also exist
 (see, e.g., \cite{Chen-BWASA-2006josab}), 
 but usually they are much less important.
The $\chi^{(3)}$ case has already been extensively addressed in the literature
 \cite{Rosen-1965pr,Flesch-PM-1996prl,Gilles-MV-1999pre,Kinsler-RTN-2007pre}.
For a $\chi^{(2)}$ nonlinearity on its own,
 eqn.~(\ref{eq:moc_n}) indicates that  
 the shocking distance depends primarily on the inverse field gradient so that
 $S_2 \propto \textrm{Min} \left| {dE}/{dt} \right|^{-1}$.

A profile for a (nearly shocked) $\chi^{(2)}$ CSS pulse
 in a dispersionless medium
 can be seen in fig.~\ref{fig:profiles}(a), 
 with its characteristic sawtooth profile.
It exhibits two distinct regions; 
 one containing a ramp-like smooth progression
 from negative to positive field values
 taking up the majority of the cycle, 
 and the second with a rapid positive to negative transition
 over the remaining small fraction of the cycle.
If we plotted the wave form for a continuous wave field, 
 rather than a pulse, 
 we would see the field evolving 
 towards a sawtooth-like profile
 (e.g. as seen in \cite{Kinsler-2007josab}).

\begin{figure}[htb]
\centering
\includegraphics[angle=-90,width=0.68\columnwidth]{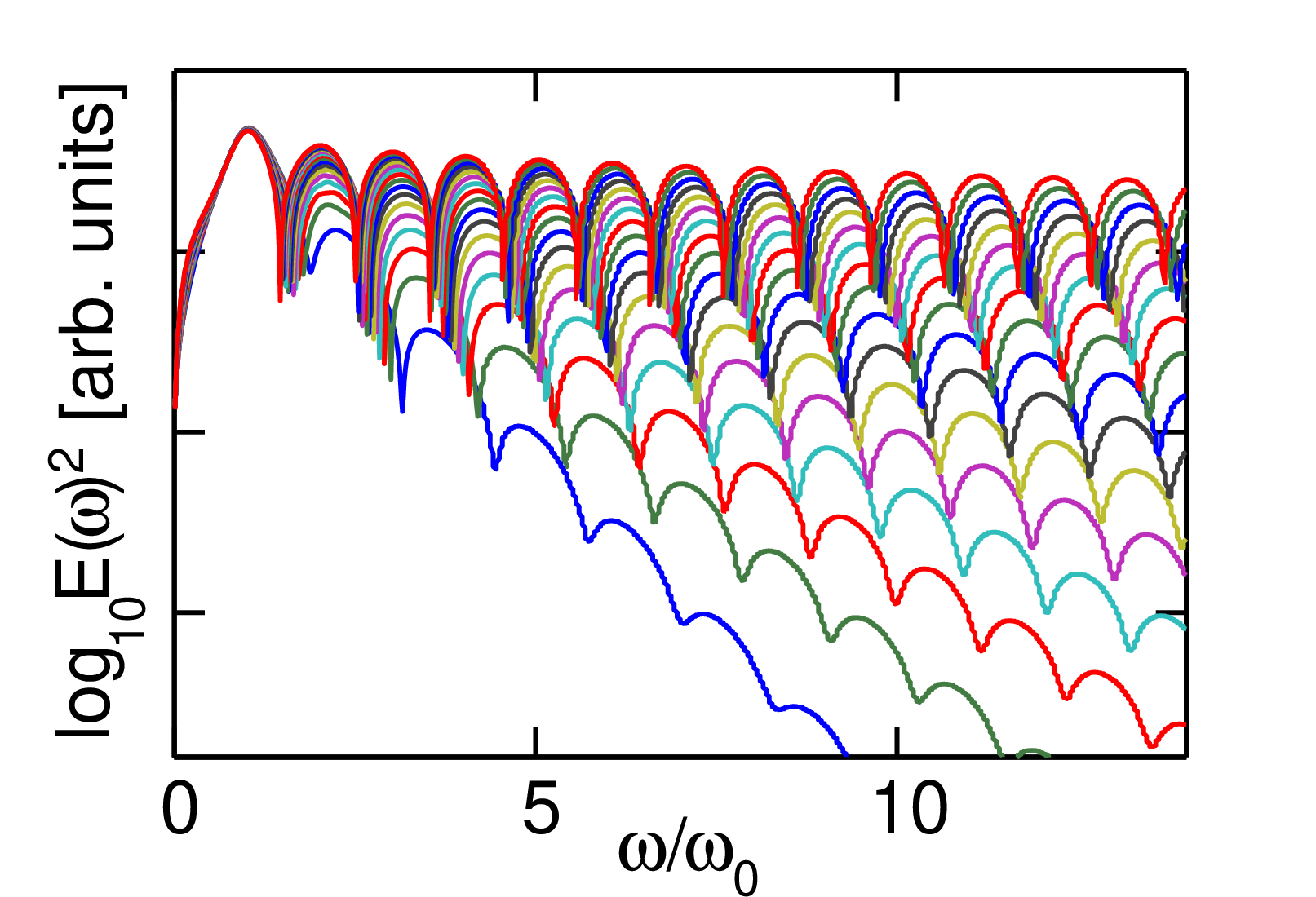}
\caption{
(Color online)
Logarithmic plot of $E(\omega)^{2}$ every half micron, 
 approaching $S_2$ in a $\chi^{(2)}$ material. 
The spectra flatten out as increasing amounts of ever higher-order harmonics
 are generated as the pulse approaches the shock.
The pulse and material parameters are as for fig.~\ref{fig:profiles}.
Each ordinate division represents five orders of magnitude.  
} \label{fig:shock2}
\end{figure}

The build-up of harmonic content
 during the reshaping and steepening of the time domain profiles
 is vividly illustrated in fig.~\ref{fig:shock2}.
Note that the spectra differ from those produced under $\chi^{(3)}$ 
 self-steepening where only odd harmonics are produced. 
The nature of the harmonic cascade in fig.~\ref{fig:shock2} indicates that, 
 if it were possible to achieve strong self-steepening experimentally,
 this would in itself be a route to HHG.  
We will return to this point in the next section.

\begin{figure}[htb]
\centering
\includegraphics[angle=-90,width=0.68\columnwidth]{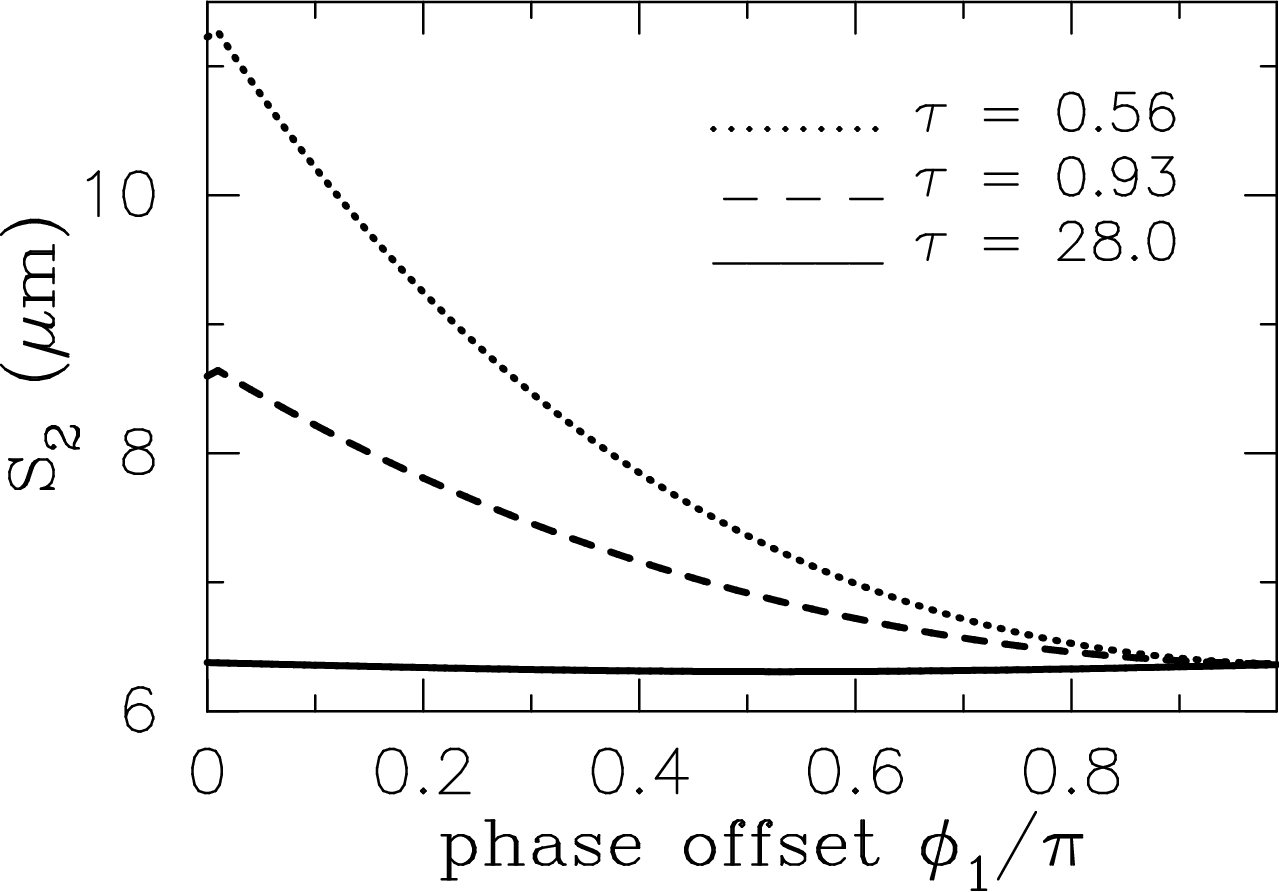}
\caption{
MOC shocking distances $S_2$ as a function of CEP $\phi$,
 allowing for different pulse widths. 
In the long pulse ($\tau=28$) case, 
 a shallow minimum centered around $\phi_1=\pi/2$ is just visible, 
 caused by the term in the square root
 of eqn.~(\ref{eq:moc_n}).
Apart from the varying pulse length, 
 the pulse and material parameters are as for fig.~\ref{fig:profiles}.
} 
\label{fig:lengthsandwidths}
\end{figure}

As for the $\chi^{(3)}$ case \cite{Kinsler-RTN-2007pre},
 the $\chi^{(2)}$ process is also sensitive to the pulse length 
 and the carrier envelope phase (CEP) offset; 
 these parameters are also important when driving HHG.
Fig.~\ref{fig:profiles}(a)
 shows that for a CEP of $\phi_1=0$,
 the carrier has a positive gradient at the center of the pulse. However, 
 steepening occurs for negative gradients in the $\chi^{(2)}$ case,
 and hence the high gradients occur near the zeros half a cycle away on either 
 side.  Since these are reduced by envelope fall-off, 
 $S_2$ increases as the pulse width is reduced for $\phi_1=0$.
In contrast, 
 when $\phi_1=\pi$, 
 steepening occurs at the center of the pulse, 
 and $S_2$ is then independent of pulse duration.  
The dependence of $S_{2}$ on CEP for different pulse widths
 is shown in Fig.~\ref{fig:lengthsandwidths}.
This can be compared with the more complicated CEP dependence
 seen in fig.~4 of \cite{Kinsler-RTN-2007pre},
 where $S_3 \propto \textrm{Min} \left| {dE^2}/{dt} \right|^{-1}$.

%
\section{Generating CSS pulses}\label{S-practical}

The theory of Section \ref{S-shocking} 
 was based on the ideal case of a dispersionless medium with 
 instantaneous nonlinearity.  
The question of whether sufficiently self-steepened pulses
 could be generated in {\em real}
 nonlinear materials has been addressed
 for the $\chi^{(3)}$ case in \cite{Flesch-PM-1996prl,Gilles-MV-1999pre}, 
 where considerations such as linear dispersion \cite{Kinsler-RTN-2007pre}
 and the strength and response time of the nonlinearities were evaluated. 
Similar issues arise in the $\chi^{(2)}$ case, 
 and so we will not repeat them here. 
We note that for many $\chi^{(2)}$ materials, 
 the polarization of the generated harmonics is different
 from that of the driving field, 
 bringing in the further complication of birefringence.
However, materials do exist that support processes where the polarizations
 are the same (e.g. $e+e\rightarrow e$ in lithium niobate \cite{dimitriev}), 
 which could (in principle) generate a harmonic cascade 
 in a single polarization.

In practice, the limitations imposed by dispersion and material damage 
 mean that generating strongly self-steepened pulses
 would be an extremely challenging task, 
 although some success might be achieved
 if relatively weak self-steepening were sufficient.
 We therefore consider the possibility of 
 {\em synthesizing} CSS-like pulses 
 by combining a sub-set of harmonic components
 using the phase offsets and amplitudes
 obtained from numerically-generated CSS profiles.
It turns out to be possible to approximate $\chi^{(2)}$ and $\chi^{(3)}$ 
 CSS-like wave forms using field strengths 
 and phases for the $n$-th harmonic ($n>1$)
 based on the formulae
~
\begin{eqnarray}
  E_n 
&=&
   F^{-\sqrt{n-1}} ~ E_1
,
\label{eqn-synth-E}
\\
  \phi_n
&=&
  n \phi_1
 +
  \left(
    n-1
  \right) 
  \pi/N
.
\label{eqn-synth-phi}
\end{eqnarray}
where $n(>1)$ is the harmonic order, 
 $N=2$ or $3$ for the $\chi^{(2)}$ and $\chi^{(3)}$ cases respectively, 
 and $F$ is a fitting factor.
With $F=4$, 
 these equations give a reasonable match to the CSS wave form 
 with the spectrum fourth from the top in fig.~\ref{fig:shock2},
 which is close to shocking.  
We note that higher-order harmonics comprise only a small fraction
 of the total CSS pulse energy.
However, 
 only a small number of harmonics would be used in a synthesized pulse. 
Quite apart from the experimental complexity
 involved in combining a large number of harmonics, 
 the HHG experiment would be pointless if the high-order harmonics
 were available already in the driving pulse!  
As we will show in Section \ref{S-HHG}, 
 sufficient gradient enhancement can be achieved
 with a driving pulse containing only two or three 
 additional harmonic components.

%
\section{HHG and CSS pulses}\label{S-HHG}

As described in the Introduction, 
 the bursts of XUV radiation produced in the electron
 ionization-recollision process that occurs when an intense
 few-cycle laser pulse interacts with an atom, 
 contain a wide range of high-order harmonics.
HHG experiments aiming to produce isolated attosecond bursts
 are typically based on few-cycle driving pulses, 
 but a number of sophisticated techniques have also been deployed, 
 including the use of two-color driving fields
 \cite{Pfeifer-GANL-2006ol,Liu-KSW-2006pra,Oishi-KSKM-2006oe,Merdji-ABCCPJNL-2007ol}, 
 polarization-gating 
 \cite{Budil-SPL-1993pra,Corkum-BI-1994ol,Sola-MECSPVBCSVSN-2006np,Sansone-BCVAFPVAVSDN-2006s}, 
 and chirp control
 \cite{Ganeev-SRBK-2007pra}.

A typical HHG spectrum from an atomic gas
 consists of a broad plateau of harmonics extending to high orders, 
 which falls off rapidly above a cut-off energy $\mathscr{E}$.  
For monochromatic driving fields 
 \cite{Krause-SK-1992prl,Schafer-YDK-1993prl,Corkum-1993prl},
~
\begin{eqnarray}
  \mathscr{E}
&=&
 I_P + 3.17U_P
,
\label{eq:E_mono}
\end{eqnarray}
where $I_P$ is the ionization potential of the atom,
 $U_P=E_1^2/(2m_e\omega)^2$ is the ponderomotive potential of the laser field,
 and $E_1$ is the maximum field strength. 
The atom used for all the simulations in this paper was neon, 
 which has a $U_P$ of $21.6$eV.
Eqn.~(\ref{eq:E_mono}) suggests that $\mathscr{E}$ can be increased
 either by increasing $E_1$ 
 or by using an atomic species with a higher $I_P$.  
However, 
 raising $E_1$ also enhances the ionization rate, 
 and this in turn can deplete the ground state, 
 causing defocusing of the laser beam and a reduction in HHG.

\begin{figure}[htb]
\includegraphics[angle=-90,width=0.75\columnwidth]{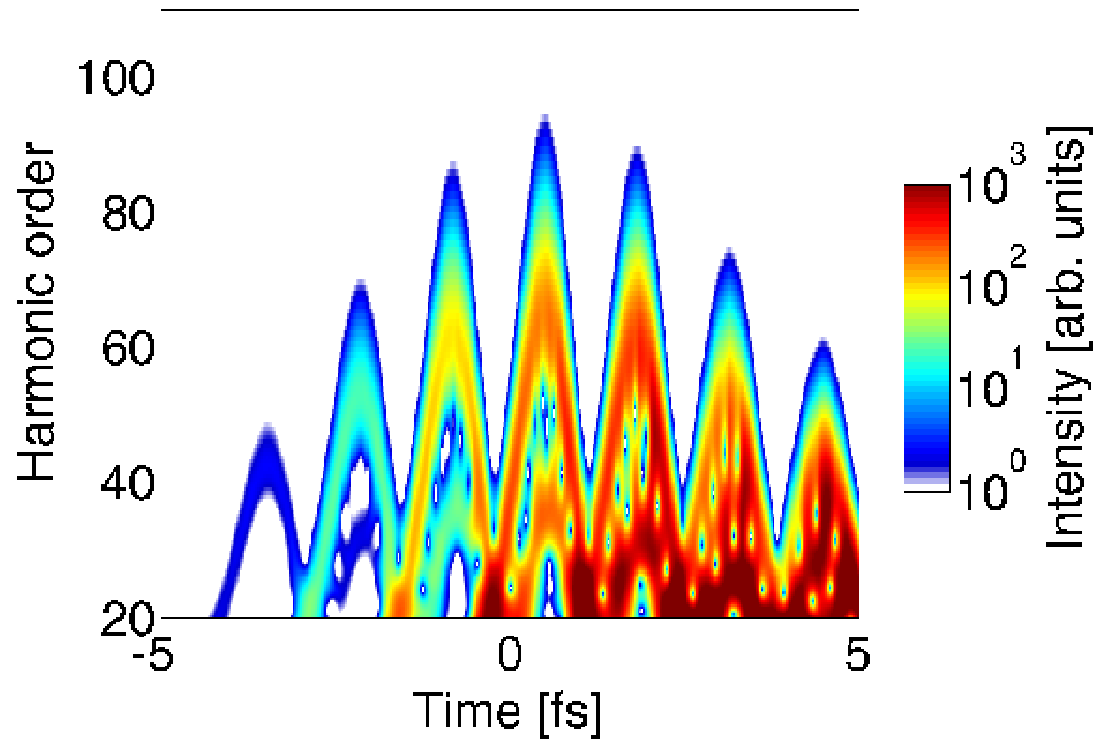}
\caption{
\label{fig:waves-normal-new}
(Color online)
Wavelet transform of the HHG
 from a normal driving pulse 
 with $\lambda = $800 nm,
 a 8-fs FWHM
 with peak intensity $5 \times 10^{14}$ W/cm$^2$.
Since the driving pulse retains inversion symmetry, 
 high-order harmonics with a similar distribution
 are generated every half-cycle.
Note that the centre of the pulse produces the bulk of the HHG,
 which [following eqn.~(\ref{eq:E_mono})] depends on $E^2$.
The 1D TDSE model of HHG
 was used to generate this figure. 
}
\end{figure}

For few-cycle driving pulses, 
 the parameters naturally change from one half-cycle to the next.  
This feature is highlighted in Fig.~\ref{fig:waves-normal-new}, 
 where a wavelet transform displays the variation
 of the harmonic spectrum for an 8-fs driving pulse, 
 calculated using the 1D time-dependent Schr\"odinger equation (TDSE) model
 \cite{Chipperfield-KTM-2006oc}. 
For pulsed (or multi-color) excitation, 
 the strong field approximation (SFA)
 \cite{Lewenstein-BIYLC-1994pra}
predicts the spectral cut-off energy to be the maximum 
 value of the function \cite{Haworth-CRKMT-2007np}
~
\begin{eqnarray}
  \mathscr{E}(t_i,t_r)
&=&
 I_P + \frac{1}{2} \left[ e A(t_r) + p(t_i,t_r)\right]^2
,
\label{E_SFA}
\end{eqnarray}
where $t_i$ is the ionization time, 
 $t_r$ the recollision time,
 $A(t)$ is the vector potential, 
 and $p$ is the \emph{asymptotic} momentum of the electron.
Eqn.~(\ref{E_SFA}) reproduces eqn.~(\ref{eq:E_mono})
 in the case of a monochromatic field. 
For electrons that 
 both ionize and recombine at the position of the core,
~
\begin{eqnarray}
  p(t_i,t_r) 
&=& 
 - 
  \frac{e}
       {t_r-t_i} 
  \int_{t_i}^{t_r} A(t) dt
.
\label{ptitf}
\end{eqnarray}

Although a general interpretation of eqn.~(\ref{E_SFA}) is complicated,
 the electron trajectories that are of most importance for HHG 
 are those for which the initial momentum of the electron
 ($p(t_i,t_r) + e A(t_i)$) is close to zero.
In this case,
 $p(t_i,t_r)=-eA(t_i)$,
 so that eqn.~(\ref{E_SFA}) simplifies to 
 $\mathscr{E}=I_P+[eA(t_r)-eA(t_i)]^2/2$.
Usually $A(t_i)$ is small, 
 so it is broadly true 
 to say that $\mathscr{E}$ depends on the peak value of $A(t_r)$.
And since the second time derivative of $A$ and, 
 concomitantly, 
 the first time derivative of $E$ tend to follow $A$,
 the implication is that maximizing the \emph{gradient}
 of the field will raise the allowed peak value of $A(t_r)$.
Since CSS-like pulses have strongly steepened field gradients, 
 they should therefore produce HHG
 with a raised cut-off energy $\mathscr{E}$
 and an increased number of harmonics.

%
\subsection{HHG using synthesized driving pulses}\label{S-HHG-synth}

To create a two-color driving field, 
 a second harmonic component is added so that 
~
\begin{eqnarray}
E(t)
&=&
E_1 \cos(\omega t + \phi_1) + E_2 \cos (2\omega t + \phi_2)
,
\label{twocolor}
\end{eqnarray} 
where we will assume that $\phi_1 = 0$ for simplicity.
The addition of the second harmonic term
 causes inversion symmetry of the profile to be lost.  
If $\phi_2 = 0$, the positive and negative lobes of the field
 have different peak amplitudes, 
 leading to different levels of tunnel
 ionization on alternate half cycles, 
 and associated variations in the intensity
 and cut-off of the HHG spectrum.
On the other hand, 
 when $\phi_2 = \pi/2$, 
 it is the positive and negative {\em slopes} of $E$ that differ, 
 along with the positive and negative excursions of $A$ and ${d^2A}/{dt^2}$.
Once again, 
 the effect is to maximize $\mathscr{E}$ every other half-cycle 
 \cite{Pfeifer-GANL-2006ol},
 but the effect is now substantially stronger than in the $\phi_2=0$ case.

\begin{figure}[htb]
\includegraphics[angle=-90,width=0.75\columnwidth]{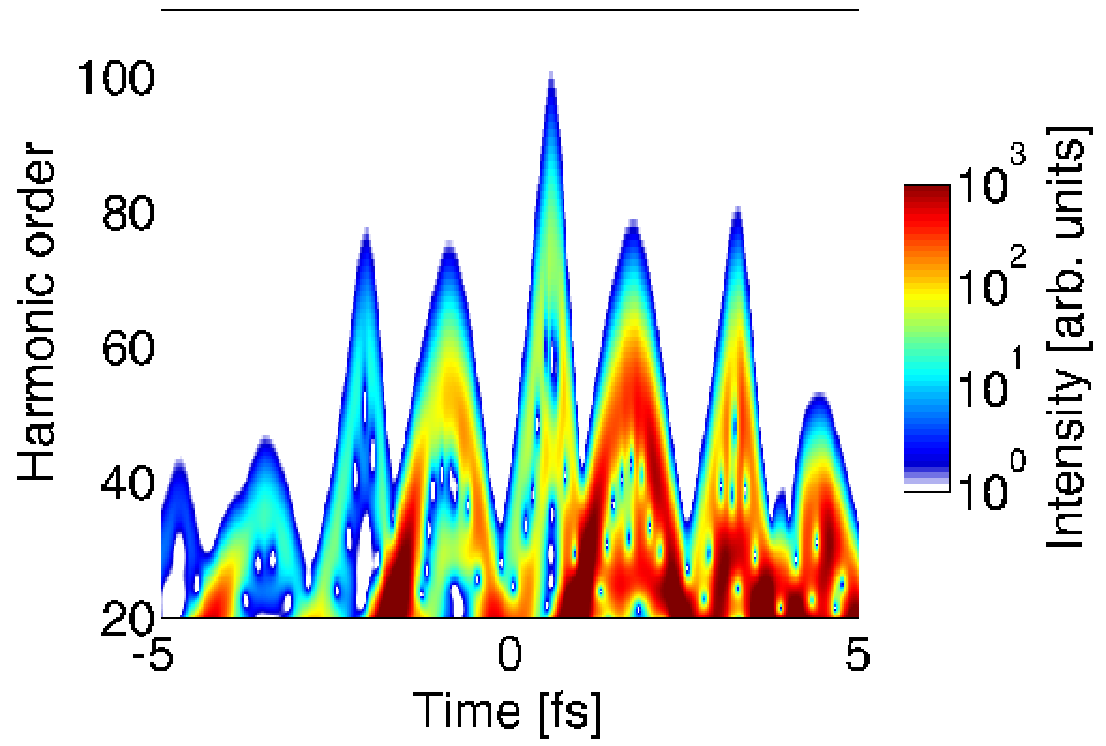}
\caption{
\label{fig:waves-chi2-new}
(Color online)
Wavelet transform of the HHG 
 from a synthesized $\chi^{(2)}$ CSS-like driving pulse using 
 amplitudes and phases
 from eqns.~(\ref{eqn-synth-E}) and (\ref{eqn-synth-phi}) 
 with $F=4$ and spectral contributions above the sixth harmonic removed.
The energy and other relevant characteristics of the driving pulse are
 same as for fig.~\ref{fig:waves-normal-new}.
It clearly shows the alternating emission from 
 each half cycle of the steepened pulse.
The 1D TDSE model of the HHG process was used to generate this figure. 
}
\end{figure}

More complex fields can be synthesized by adding further harmonics,
 with each additional component providing
 a significant boost to the field gradient.
However, 
 as we shall see in subsection \ref{S-HHG-extra} below, 
 the benefits are eventually
 subject to the law of diminishing returns.  
Pulse profiles containing
 even as well as odd harmonics invariably exhibit characteristics
 that alternate from one half-cycle to the next.  
This feature is clearly evident in Fig.~\ref{fig:waves-chi2-new}
 where the wavelet transform of the HHG generated
 by a $\chi^{(2)}$ CSS-like pulse spectrally truncated above
 the sixth harmonic is displayed; details are given in the caption. 
The figure should be compared with Fig.~\ref{fig:waves-normal-new}.
If a high-pass filter 
 (such as a molybdenum-silicon multilayer mirror)
 is applied to the signal of Fig.~\ref{fig:waves-chi2-new},
 the lower peaks can be removed, 
 leaving the highest peak(s) in greater isolation.
As such, 
 these filters are useful for the generation of a single XUV burst,
 or when creating burst trains
 for stroboscopic imaging \cite{Remetter-JMVNLGKKLSVL-2006np}.

%
\subsection{Phasing of the harmonic components}\label{S-HHG-phase}

\begin{figure}[htb]
\includegraphics[angle=-90,width=0.75\columnwidth]{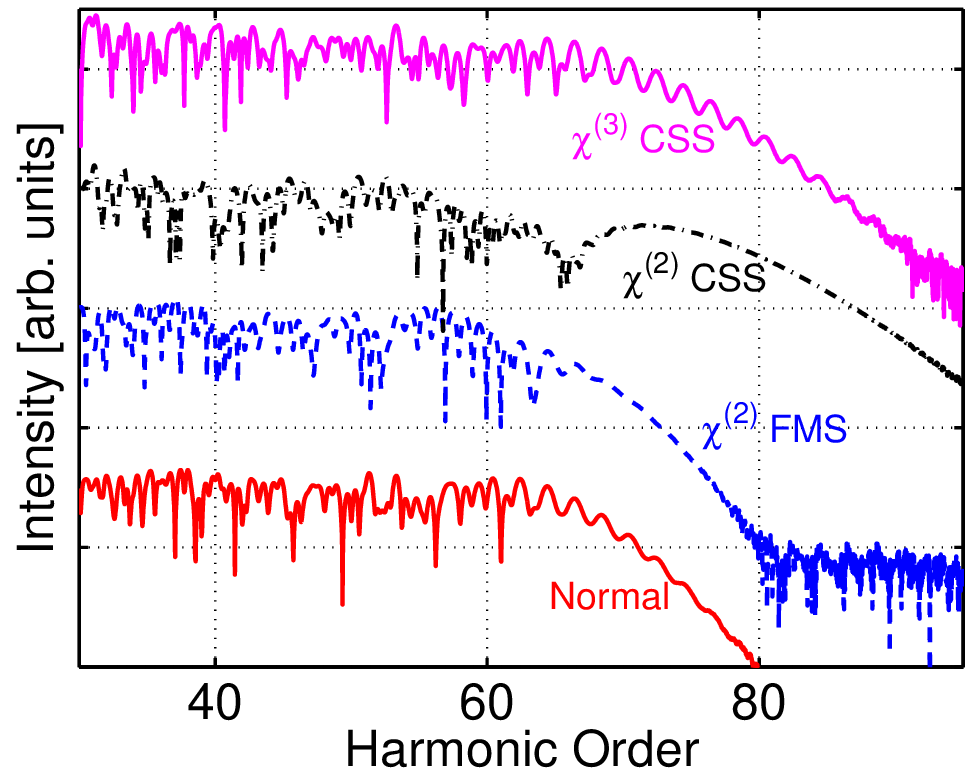}
\caption{
\label{fig:attosec-spect-new}
(Color online)
HHG spectra for different driving pulses, 
 offset vertically for clarity.
Each ordinate division represents four orders of magnitude.
The pulse energy and other relevant characteristics
 are the same as for fig.~\ref{fig:waves-normal-new}.
The synthesized $\chi^{(2)}$ FMS and CSS pulses
 contain second, third, and fourth harmonics; 
 the synthesized $\chi^{(3)}$ CSS pulse
 contains third and fifth harmonics.
Bottom curve to top curve:
 normal (red), 
 $\chi^{(2)}$ FMS (blue),
 $\chi^{(2)}$ CSS (black),
 and $\chi^{(3)}$ CSS (magenta).
}
\end{figure}

Increasing the number of harmonics
 naturally increases the experimental complexity, 
 and arranging for their correct phasing makes the situation
 even more complicated.  
It is therefore important to check that CSS-like pulses
 are better for HHG because of their steep field gradients,
 rather than simply because of their greater harmonic content.
Is the mere presence of many harmonics (irrespective of their phase) 
 the crucial thing, 
 or will a restricted range of harmonics, 
 correctly phased to maximize the gradient, 
 perhaps yield a better result with less experimental effort?  
To address this issue, 
 we compare HHG spectra for different types of driving pulse
 in fig.~\ref{fig:attosec-spect-new}.  
Reading from top to bottom, spectra are shown for

\begin{description}

\item[~]\textbullet ~
 a synthesized CSS-like pulse containing third and fifth harmonics 
(labeled $\chi^{(3)}$ CSS);

\item[~]\textbullet ~
 a synthesized CSS-like pulse containing second, third and fourth harmonics
(labeled $\chi^{(2)}$ CSS);

\item[~]\textbullet ~
 a ``field-maximized spectrum'' (FMS) pulse
 containing identical spectral content to the $\chi^{(2)}$ CSS pulse, 
 but with phases $\phi_n=0$,
 to create a maximized field rather than a maximized gradient;

\item[~]\textbullet ~
 a normal quasi-monochromatic pulse similar to the one used to create
fig.~\ref{fig:waves-normal-new}.

\end{description}

The benefit of a gradient-maximized pulse
 over both the normal and FMS pulses
 is demonstrated vividly in the figure.  
The highest $\mathscr{E}$ is clearly obtained for the
 $\chi^{(2)}$ CSS pulse.
Further, 
 a comparison of $\chi^{(2)}$ CSS,
 $\chi^{(2)}$ FMS, 
 and the normal spectra
 reveals that the CSS pulses have the most gradual cut-off.
The conclusion is that optimum performance results 
 from the character of the CSS profiles 
 and not merely their harmonic content.

\begin{figure}[htb]
\centering
\includegraphics[angle=-90,width=0.75\columnwidth]{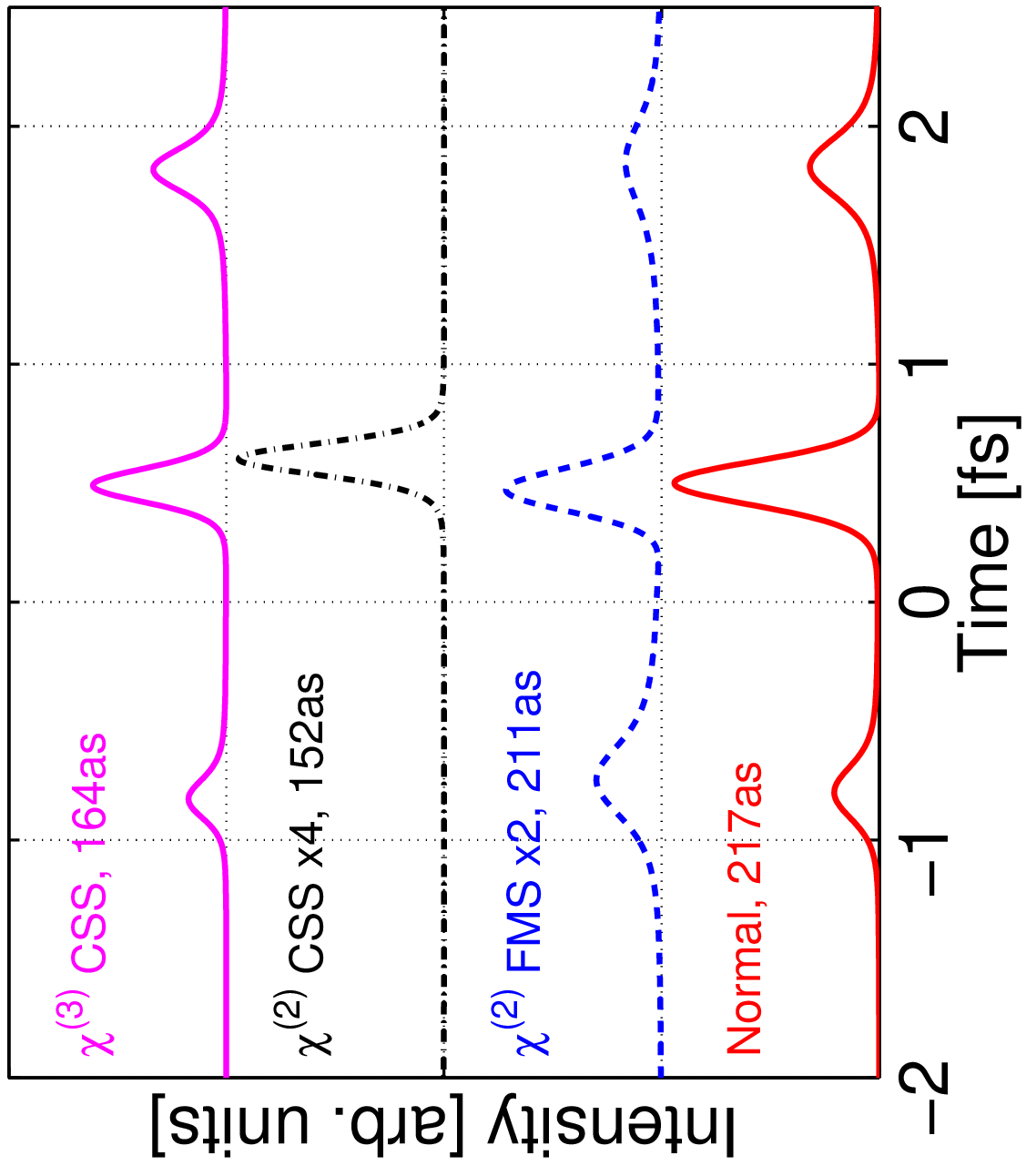}
\caption{\label{fig-attosec}
(Color online)
XUV burst intensity profiles and FWHMs
 resulting from high-pass-filtered 
 HHG spectra for the various CSS driving pulses 
 seen in fig.~\ref{fig:attosec-spect-new}.
Bottom curve to top curve:
 normal (red), 
 $\chi^{(2)}$ FMS (blue),
 $\chi^{(2)}$ CSS (black),
 and $\chi^{(3)}$ CSS (magenta).
The curves have been individually scaled to aid visibility.
}
\end{figure}

Fig.~\ref{fig-attosec}
 shows the temporal profiles of attosecond XUV bursts obtained by
 appropriate filtering of the four HHG spectra of  
 fig.~\ref{fig:attosec-spect-new}.
In each case, the filtering was adjusted to  
 minimize the full width at one {\em quarter} maximum, 
 a procedure that created the cleanest pulses. 
It is clear that a $\chi^{(2)}$ CSS driving pulse
 produces the shortest and most isolated XUV bursts, 
 although the intensity is lower than in the $\chi^{(2)}$ FMS case
 where there is one strong and one weak burst per cycle.
The $\chi^{(3)}$ CSS pulse
 produces \emph{two} short XUV bursts per cycle, 
 with a relatively high intensity. 
We note that the oscillations in the tails 
 of the $\chi^{(3)}$ CSS spectra in fig.~\ref{fig:attosec-spect-new}
 are associated with the presence of more than one XUV burst, 
 as seen in Fig.~\ref{fig-attosec}.

%
\subsection{Optimum number of harmonics}\label{S-HHG-extra}

As mentioned earlier, 
 moving from standard single-color to two-color pumping 
 adds experimental complexity, 
 and the situation will naturally be exacerbated
 if additional harmonics are included.  
It is therefore important to study how effectively the 
 extra experimental investment
 is reflected in improved HHG performance.

\begin{figure}[htb]
\centering
\includegraphics[angle=-90,width=0.75\columnwidth]{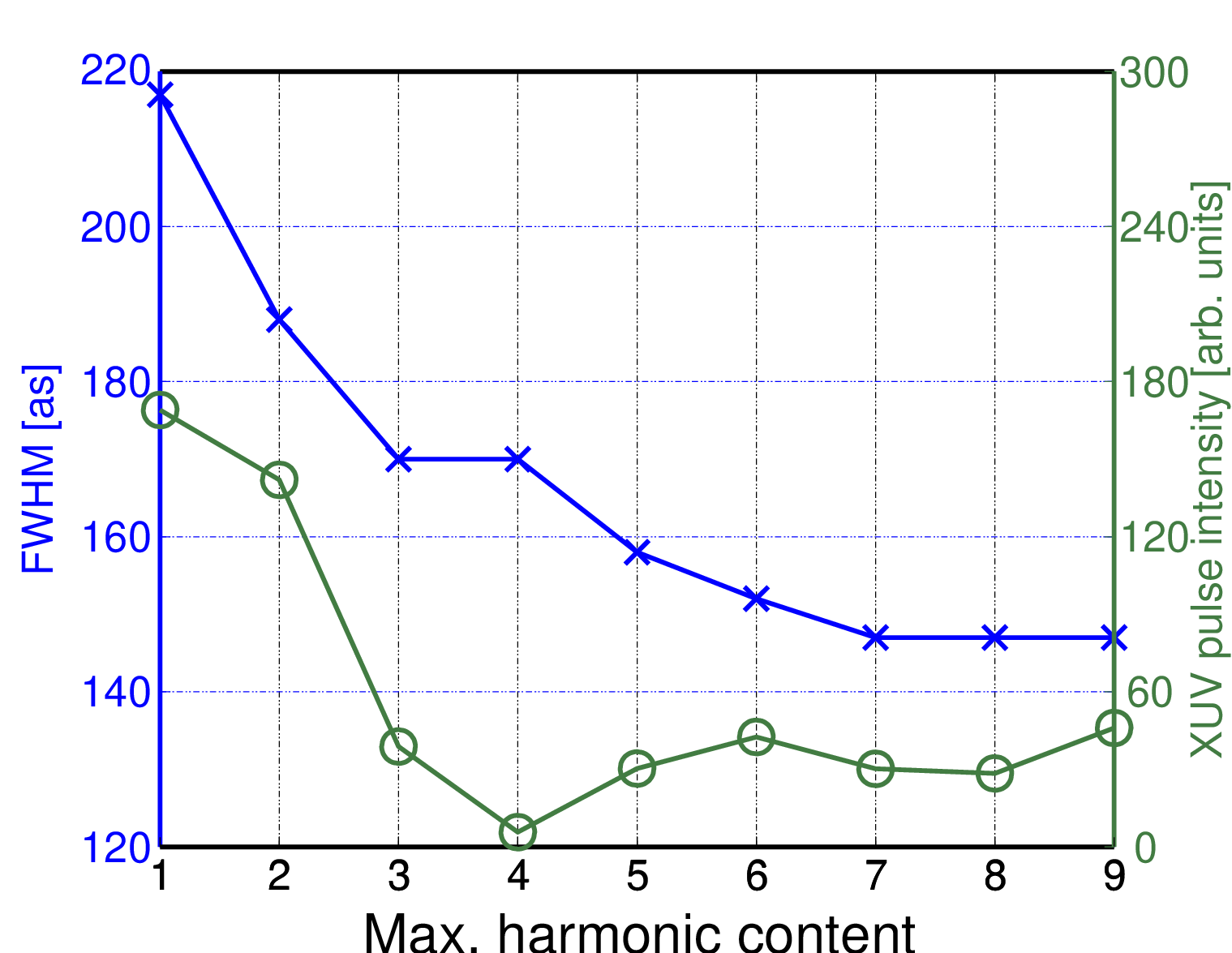}
\caption{\label{fig:synthesis2}
(Color online)
XUV burst durations and intensities resulting from high-pass-filtered 
 HHG spectra for synthesized $\chi^{(2)}$ CSS-like driving pulses,
 as a function of the maximum harmonic component included.  
The burst durations are denoted with $\times$ 
 and use the left-hand scale; 
 the intensities are denoted with $\circ$ 
 and use the right-hand scale. 
The 1D TDSE model was used.
}
\end{figure}

\begin{figure}[htb]
\centering
\includegraphics[angle=-90,width=0.75\columnwidth]{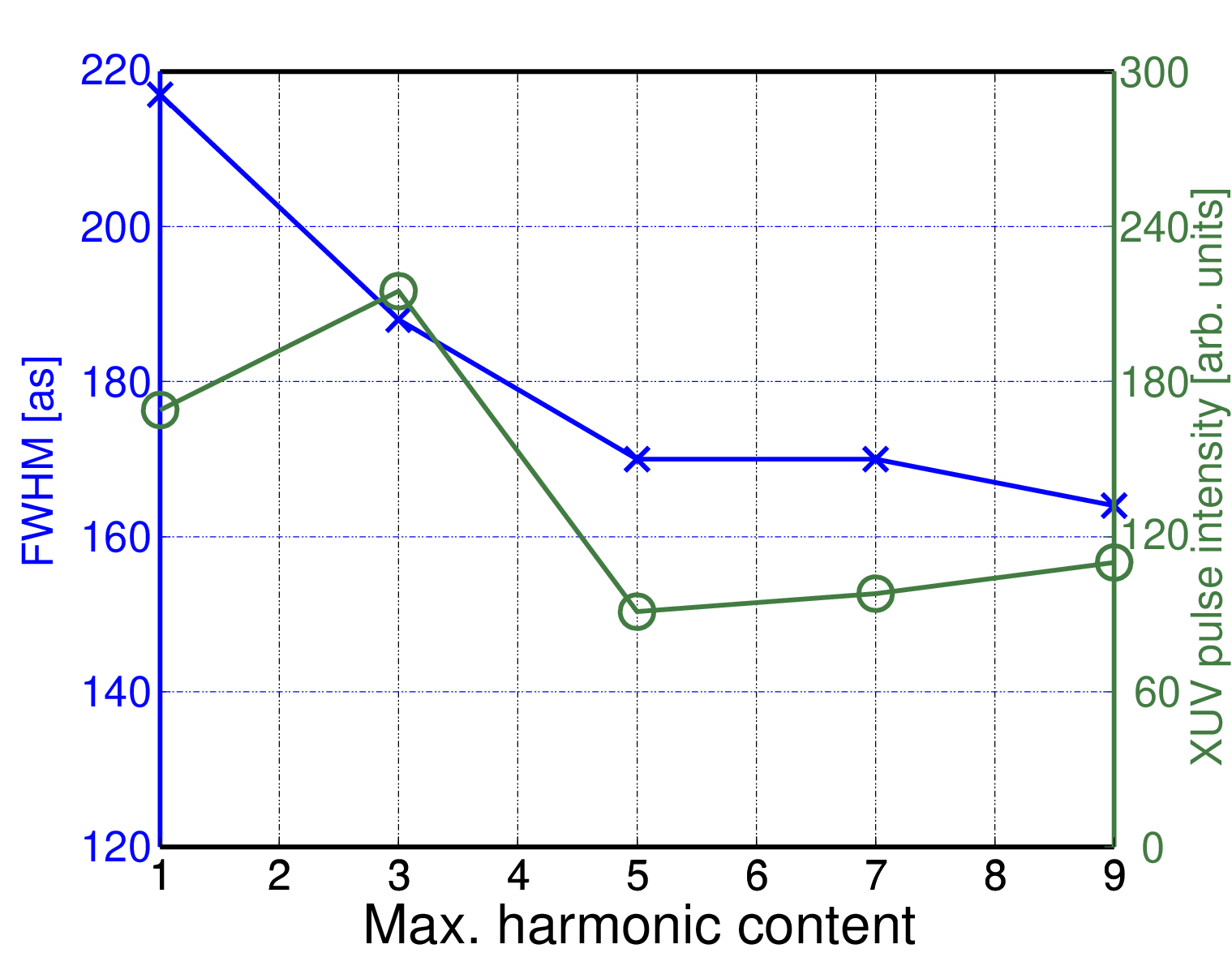}
\caption{\label{fig:synthesis3}
(Color online)
XUV burst durations and intensities resulting from high-pass-filtered 
 HHG spectra for synthesized $\chi^{(3)}$ CSS-like driving pulses,
 as a function of the maximum harmonic component included.  
The burst durations are denoted with $\times$ 
 and use the left-hand scale; 
 the intensities are denoted with $\circ$ 
 and use the right-hand scale. 
The 1D TDSE model was used.
}
\end{figure}

Figs.~\ref{fig:synthesis2} and \ref{fig:synthesis3}
 demonstrate what happens when HHG is driven
 by $\chi^{(2)}$ and $\chi^{(3)}$ CSS-like pulses
 as the number of harmonic components is increased. 
In each case, 
 there is clearly a strong initial reduction
 in the burst durations as extra harmonics are added. 
The FWHM decreases roughly linearly up to about the fifth harmonic, 
 but the graphs flatten out thereafter. 
The intensity of the bursts drops off rapidly too, 
 although this is not necessarily a serious consideration; 
 after all, 
 much intensity is also lost in 
 polarization-gating schemes.
Different filtering strategies (e.g. allowing lower frequencies through) 
 can provide shorter XUV bursts than shown in fig.~\ref{fig:synthesis2},
 but this usually results in stronger satellite pulses.

%
\subsection{Optimum field profile}\label{S-HHG-profile}

\begin{figure}[htb]
\centering
\includegraphics[angle=-90,width=0.75\columnwidth]{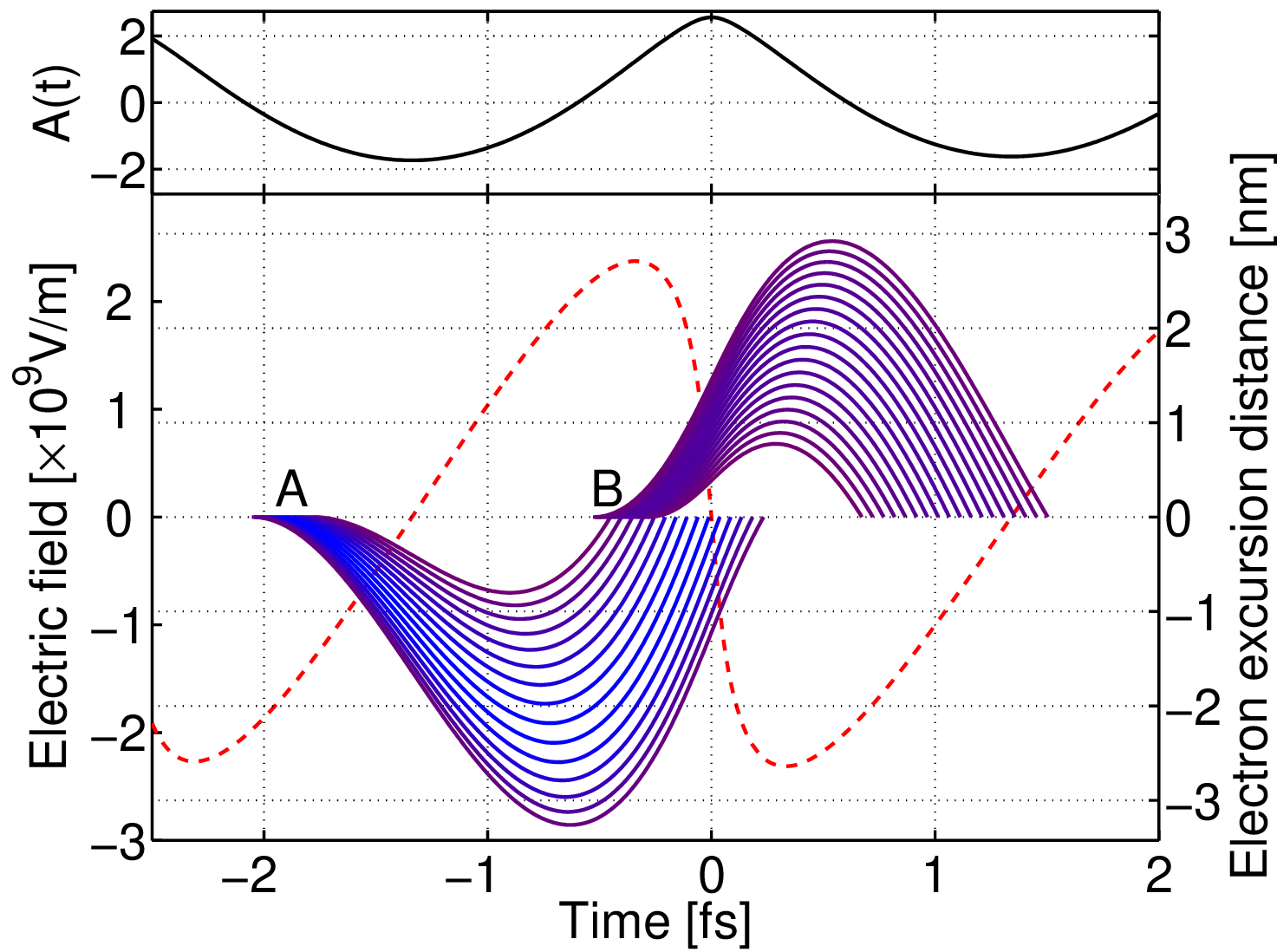}
\caption{
\label{fig:chi2schematic}
(Color online)
Upper frame:
 the vector potential $A(t)$ corresponding to the field profile
 shown in the lower frame.
Note how the peaks in $A(t)$ correspond to the regions
 of high field gradient.
Lower frame:
 the trajectories of {\em only} those high-energy ionized electrons 
 that recollide with the nucleus to emit high-order harmonic radiation, 
 calculated using the SFA model
 \cite{Lewenstein-BIYLC-1994pra}.
The dashed (red) line shows the CSS field profile, 
 in this case for the centre of a three-cycle pulse with peak intensity
 $5 \times 10^{14}$W/cm$^2$.
The CSS-enhanced trajectories (starting at A, on the left)
 recollide over a short period of time at high field gradients,  
 giving rise to a brief XUV emission with larger cut-off energy.
The latter set of recolliding trajectories (starting B, right)
 have a lower energy,
 and recollide over a longer time interval at low field gradients.
}
\end{figure}

The fact that $\chi^{(2)}$ CSS pulses succeed in maximizing
 the HHG cut-off energy $\mathscr{E}$ at a given pulse energy
 is due to the way in which the particular electric field profile
 controls the motion of the ionized electrons.  
As shown in fig.~\ref{fig:chi2schematic}, 
 the outward journey and initial deceleration of the electrons
 in group A are determined by the shallow ramp of the field
 as it rises gently from negative to positive.
The high return acceleration is managed by the high-field region, 
 and the subsequent steep field decrease
 allows the electrons to recollide with the core within minimal deceleration,
 whilst simultaneously resetting the field amplitude
 so that the process can repeat.
The overall effect is to create trajectories
 that minimize the distance traveled, 
 while ensuring the highest possible recollision energies
 along with a high $\mathscr{E}$.
The high-energy recollisions take place
 largely during the brief high-gradient part of the field profile; 
 and give rise to the narrow peaks 
 in fig.~\ref{fig:waves-chi2-new}, 
 which then become the short duration XUV bursts 
 of fig.~\ref{fig-attosec} after high-pass filtering.

A genetic algorithm \cite{Forrest-1993sci} used in combination
 with classical simulations of electron trajectories, 
 has confirmed that the sawtooth-like $\chi^{(2)}$ CSS profiles 
 are uniquely efficient for HHG. 
Using parametrized field profiles,
 we sought to optimize the recollision energy
 while holding both the pulse energy per cycle
 and the periodicity fixed.  
The algorithm consistently predicted that the optimum wave form 
 was a linear ramp
 starting at a field $-E_{max}/2$ and increasing to $E_{max}$.  
This corresponds to the last three quarter-cycles 
 of a sawtooth wave form; 
 indeed in fig.~\ref{fig:chi2schematic} we can see that 
 SFA predictions for the CSS-enhanced trajectories
 agree with the classical picture and 
 start at about a quarter cycle in.
Adding a DC bias or sufficiently low-frequency field 
 to a $\chi^{(2)}$ CSS-like profile 
 might therefore make fuller use
 of the whole ramp section of the wave form, 
 further increasing $\mathscr{E}$.

A second family of ionization/ recollision trajectories 
 (group B in fig.~\ref{fig:chi2schematic})
 are initiated during the high-field region of the optical cycle. 
However, 
 these recollisions occur in a region of lower field, 
 produce lower recollision energies, 
 and are spread over a longer time interval.
They lead to more intense but longer HHG emission,
 with a low $\mathscr{E}$,
 represented by the broader and lower peaks in fig.~\ref{fig:waves-chi2-new}.

Returning to 
 figs.~\ref{fig:synthesis2} and \ref{fig:synthesis3}, 
 we see that the change in burst duration (or intensity)
 for each additional harmonic component
 is not necessarily straightforward or monotonic. 
This serves to illustrate that 
 that HHG dynamics are complicated, 
 and that the general principle of maximizing the field gradient
 could be optimized by further use of a genetic algorithm.  
Indeed, 
 any HHG model could be optimized for a specific outcome --
 e.g. for the shortest-duration XUV burst.  
It would also allow the potential benefits of spectral inputs 
 other than the harmonic cascade used here to be evaluated.

%
\section{Conclusion}\label{S-conclusion}

We have shown how to optimize HHG using gradient-gating of the 
 interaction, 
 and that CSS pulses (or in practice their synthesized counterparts)
 are the most efficient way of providing 
 the localized steep field gradients needed.
An important conclusion is that the gradient-gated HHG enhancement
 can be achieved using as few as three or four 
 correctly phased extra harmonic components.
The study was assisted by our generalized theory of 
 optical carrier-wave self-steepening.
By describing wave forms generated by a $\chi^{(2)}$ nonlinear interaction, 
 the theory showed how 
 their advantageous symmetry properties could be used to generate shorter, 
 more isolated XUV bursts.
By using a genetic algorithm with selection
 based on the recollision energy of classical electrons, 
 we confirmed that indeed the sawtooth-like $\chi^{(2)}$ CSS pulses
 do have an optimal form that increased the HHG cut-off energy $\mathscr{E}$
 for any given pulse energy.

In summary, 
 we have shown that the use of CSS-like driving pulses
 for HHG has the potential to generate shorter, 
 more isolated XUV bursts, 
 as well as suggesting possibilities for other HHG optimization schemes.

%
\appendix
\section{MOC theory}\label{S-MOC}

Our starting point is the pair of 1D sourceless, 
 plane polarized Maxwell's equations for a field
 propagating in the $z$ direction,
 namely:
\begin{eqnarray}
\label{eq:maxwell}
  \frac{\partial E_{x}}{\partial z}
&=&
  -\mu_{0}\frac{\partial H_{y}}{\partial t}
; ~~~~ ~~~~ 
  -\frac{\partial  H_{y}}{\partial z}
=
  \frac{\partial D_{x}}{\partial t}.
\end{eqnarray}
where $E_x$, $H_{y}$ and $D_{x}$ are the electric, 
 magnetic and electrical displacement fields respectively.  
We neglect the tensor nature of the nonlinear coefficients 
 since the polarization of the generated field 
 can be the same as that of the incoming field in some situations. 
Under these circumstances, 
 and writing $E=E_x$ for convenience,
 we have: 
~
\begin{eqnarray}
\label{eq:d-i}
  D
&=&
  \epsilon_{0}\left(E+\chi^{(1)}E+\sum_{m>1}\chi^{(m)}E^{m}\right),
\end{eqnarray}
where $\chi^{(m)}$ refers
 to the $m$-th order nonlinear susceptibility of the medium, 
 which is assumed to be instantaneous. 
This leads to a factorizable second order wave equation, 
 whose forward propagating part is
\begin{eqnarray}
\label{eq:waves}
  \left(
    \frac{\partial}{\partial t}
   -v(E)\frac{\partial}{\partial z}
  \right)E
&=&
  0
\end{eqnarray}
where the characteristic velocity $v(E)$ is,
\begin{eqnarray}
\label{eq:sigma}v(E)=\frac{c}{\sqrt{n_{0}^{2}+\sum_{m>1}m\chi^{(m)}E^{m-1}}}.
\end{eqnarray}

Using the MOC, it is possible to obtain an analytical formula
 for the shocking distance; 
 the technique is used for initial value problems
 in many fields \cite{whitham}. 
The objective is to turn a first-order partial differential equation (PDE)
 into an ordinary differential equation by changing the co-ordinates. 
In our case, 
 the second-order PDE in eqn.~(\ref{eq:waves})
 separates into two first-order PDEs. 
The characteristic curves then follow the field contours, 
 and a shock develops where characteristics converge, 
 which occurs if a later point travels faster than an earlier point. 
If the characteristics diverge, no shock develops.

If one characteristic travels slower than another, 
 they will intersect after some distance $L$, 
 since $\frac{dv}{dt}
\simeq
  \frac{v}{t}
=
  \frac{v^{2}}{L}$.
Substituting eqn.~(\ref{eq:sigma}) into this, 
 we arrive at the generalized formula for the distance 
 at which characteristics converge:
~
\begin{eqnarray}
  \label{eq:moc_n-i}
  L
&=&
 - 
    \frac{2c\sqrt{n_{0}^{2}+\displaystyle\sum_{q>1} q\chi^{(q)}E^{q-1}}}
         { \displaystyle\sum_{m>1} m \chi^{(m)}\frac{dE^{m-1}}{dt}}
.
\end{eqnarray}
The shocking distance ($S$) is now the minimum value of $L$, 
 i.e.  $S=\text{Min}|L|$.
This minimum can be easily found numerically. 


%
\section*{Acknowledgement}

The authors acknowledge financial support
 received from the Engineering and Physical Sciences Research Council (U.K.).

%

%

\end{document}